\documentclass[a4paper,11pt]{article}
\usepackage{jheppub} 
\usepackage{lineno}

\newcommand{\smartpixels}{\emph{smartpixels}\xspace}

\usepackage{graphicx} 
\usepackage{dcolumn} 
\usepackage{bm} 
\usepackage{slashed}
\usepackage[dvipsnames]{xcolor}
\usepackage{xspace}
\usepackage{amsmath}
\usepackage{amsfonts}
\usepackage{comment}
\usepackage{lmodern}
\usepackage{tabulary}
\usepackage{dsfont}
\usepackage{graphicx}
\usepackage{subcaption}
\usepackage{hyperref}
\usepackage{placeins}
\usepackage{subcaption}

\usepackage{tabularx,array}

\newcolumntype{C}{>{\centering\arraybackslash}X}
\usepackage{multirow}
\usepackage{siunitx}
\DeclareSIUnit{\electron}{e^-}
\usepackage[colorlinks=true,
urlcolor=blue,
linkcolor=blue,
citecolor=blue,
linktocpage=true,
pdfproducer=medialab,
pdfa=true,
anchorcolor=blue]{hyperref}

\usepackage{float}
\usepackage{xcolor}

\usepackage{defs}

\leftline{FERMILAB-PUB-25-0684-ETD}
\leftline{Version 2 as of \today}

\title{Characterization of a 28 nm \textit{smartpixels} ASIC With On-Chip ML for Particle Tracking Detectors} 

\author[a,c]{Benjamin Parpillon}
\author[b]{Anthony Badea}
\author[c]{Danush Shekar}
\author[a]{Cristian Gingu}
\author[a,d]{Giuseppe Di Guglielmo}
\author[a]{Tom Deline}
\author[a]{Sergey Los}
\author[a]{Adam Quinn}
\author[e,f]{Michele Ronchi}
\author[b]{Daniel Abadjiev}

\author[a]{Doug Berry}
\author[h]{Arghya Ranjan Das}
\author[g]{Jennet Dickinson}
\author[b]{Karri DiPetrillo}
\author[a,b,d]{Farah Fahim}
\author[a]{Lindsey Gray}
\author[c]{Harshul Gupta}
\author[b]{Eliza Howard}
\author[i]{David Jiang}
\author[a]{Pamela Klabbers}
\author[b]{Mira Littmann}
\author[h]{Mia Liu}
\author[j]{Petar Maksimovic}
\author[l]{Nick Manganelli}
\author[c]{Corrinne Mills}
\author[i]{Mark S. Neubauer}
\author[k]{Jannicke Pearkes}
\author[a]{Paul Rubinov}
\author[k]{Ricardo Silvestre}
\author[j]{Morris Swartz}
\author[a]{Chinar Syal}
\author[a,d]{Nhan Tran}
\author[c]{Amit Trivedi}
\author[k]{Keith Ulmer}
\author[c]{Mohammad Abrar Wadud}
\author[g]{Benjamin Weiss}
\author[c]{Jieun Yoo}
\author[b]{Eric You}

\emailAdd{bparpill@fnal.gov, badea@uchicago.edu, dsheka3@uic.edu} 

\affiliation[a]{Fermi National Accelerator Laboratory, Batavia, IL 60510, USA}
\affiliation[b]{The University of Chicago, Chicago, IL 60637, USA}
\affiliation[c]{University of Illinois Chicago, Chicago, IL, 60607, USA}
\affiliation[d]{Northwestern University, Evanston, IL 60208, USA}
\affiliation[e]{Politecnico di Milano, DEIB, Milano, 20133, Italy}
\affiliation[f]{INFN, Sezione di Milano, Milano, 20133, Italy}
\affiliation[g]{Cornell University, Ithaca, NY 14853, USA}
\affiliation[h]{Purdue University, West Lafayette, IN 47907, USA}
\affiliation[i]{University of Illinois Urbana-Champaign, Champaign, IL 61801, USA}
\affiliation[j]{Johns Hopkins University, Baltimore, MD 21218, USA}
\affiliation[k]{University of Colorado Boulder, Boulder, CO 80309, USA}
\affiliation[l]{Northeastern University, Boston, MA 02115, USA}

\abstract{
We present a 28~nm CMOS pixel readout integrated circuit implementing in-pixel analog signal processing and on-chip machine learning data filtering for particle tracking detectors. Our ASIC comprises two $32 \times 8$ pixel matrices with a pixel pitch of $25 \times 25~\mu\mathrm{m}^2$, in which each pixel integrates a charge-sensitive amplifier with synchronous auto-zero offset cancellation and a 2-bit flash ADC with programmable thresholds. Two analog front-end architectures, single-ended and differential, are implemented and characterized. Digitized pixel data are combined into row-wise projections and processed by an on-chip, fully combinational neural network classifier for data reduction. Measurements at room temperature using charge injection demonstrate an equivalent noise charge of $54.6~\mathrm{e}^{-}$ and a threshold dispersion of $\sim$78.2~\unit{\electron} at nominal bias, linear response up to several~\unit{\kilo\electron}, and stable operation at a 10~MHz clock frequency. The neural network output is compared with offline RTL predictions and agrees for $99.06\%$ of $1.5 \times 10^{5}$ test inputs.
}

\keywords{high energy physics, particle tracking, microelectronics, machine learning}

\begin{document}
\maketitle
\flushbottom

\section{Introduction}
\label{sec:intro}

Modern silicon pixel detectors at the High-Luminosity Large Hadron Collider (HL-LHC) will operate under unprecedented particle occupancies, producing data rates that exceed the practical bandwidth available for off-detector transmission. While advances in sensor technology continue to improve spatial resolution and timing performance, the increasing data volume places growing pressure on readout architectures, power consumption, and trigger systems. Future detectors therefore require new approaches capable of reducing data while preserving the physics information relevant for reconstruction.

Conventional pixel readout ASICs perform only limited local processing, typically consisting of amplification, discrimination, and zero suppression before transmitting hit information to downstream electronics~\cite{RD53BmeasGaioni,Crocv1,ITKv1} . Although effective, these approaches treat every surviving hit equally and therefore cannot exploit the rich spatial information contained within charge clusters to distinguish signal from background. Consequently, a large fraction of the transmitted data contributes little to the final physics analysis while consuming valuable bandwidth.
These limitations motivate moving more advanced processing closer to the sensor, where charge cluster information can be analyzed before the data are transmitted off detector. 
In particular, on-chip neural networks (NN) can exploit charge clusters deposited in a single detector layer from particles traversing the detector to perform data reduction at the point of acquisition.
Prior studies by the \smartpixels collaboration~\cite{smartpixels_website} established the feasibility of implementing a NN on an ASIC to perform real-time filtering of incident particles by transverse momentum~\cite{Yoo_2024, shekar2025}. Further work explores the possibility of regressing the incident track parameters and their uncertainties using the cluster shape~\cite{smartpix_regression}.

An initial prototype implementing only the analog front-end (AFE) was fabricated and characterized prior to the integration of the machine-learning circuitry, as reported in Ref.~\cite{ISCAS2023}. The subsequent \smartpixels ASIC, which integrates the AFE with the filtering algorithm, was presented in Ref.~\cite{parpillon2024smartpixelsinpixelai}. Following the tape-out of this design in a TSMC 28 nm CMOS process, this paper presents the silicon characterization of the mixed-signal pixel readout architecture. Additionally, studies of the on-chip NN algorithm are included to demonstrate end-to-end operation of the architecture rather than investigating the performance of the network itself. Accordingly, we discuss only the aspects of the model necessary to understand the hardware implementation; the training procedure and physics-relevant performance of the network will be presented in a future publication.

The remainder of this paper is organized as follows. Section~II describes the \smartpixels ASIC, including the analog front-end, synchronous digitization, and on-chip classifier. Section~III presents the data-acquisition and testing system. Section~IV reports the analog front-end characterization
and discusses the principal first-silicon limitations. Section~V evaluates the power consumption and RTL-to-silicon agreement of the on-chip neural network. Section~VI summarizes the results and outlines the planned hardware and algorithmic developments.

\section{\smartpixels ASIC}
\label{sec:asic}

\subsection{Pixel Analog Front-End and Synchronous Digitization}
\begin{figure*}[p]
    \centering
    \vspace*{\fill}

    \begin{subfigure}[t]{0.98\textwidth}
        \centering
        \includegraphics[
            width=\linewidth,
            height=0.3\textheight,
            keepaspectratio
        ]{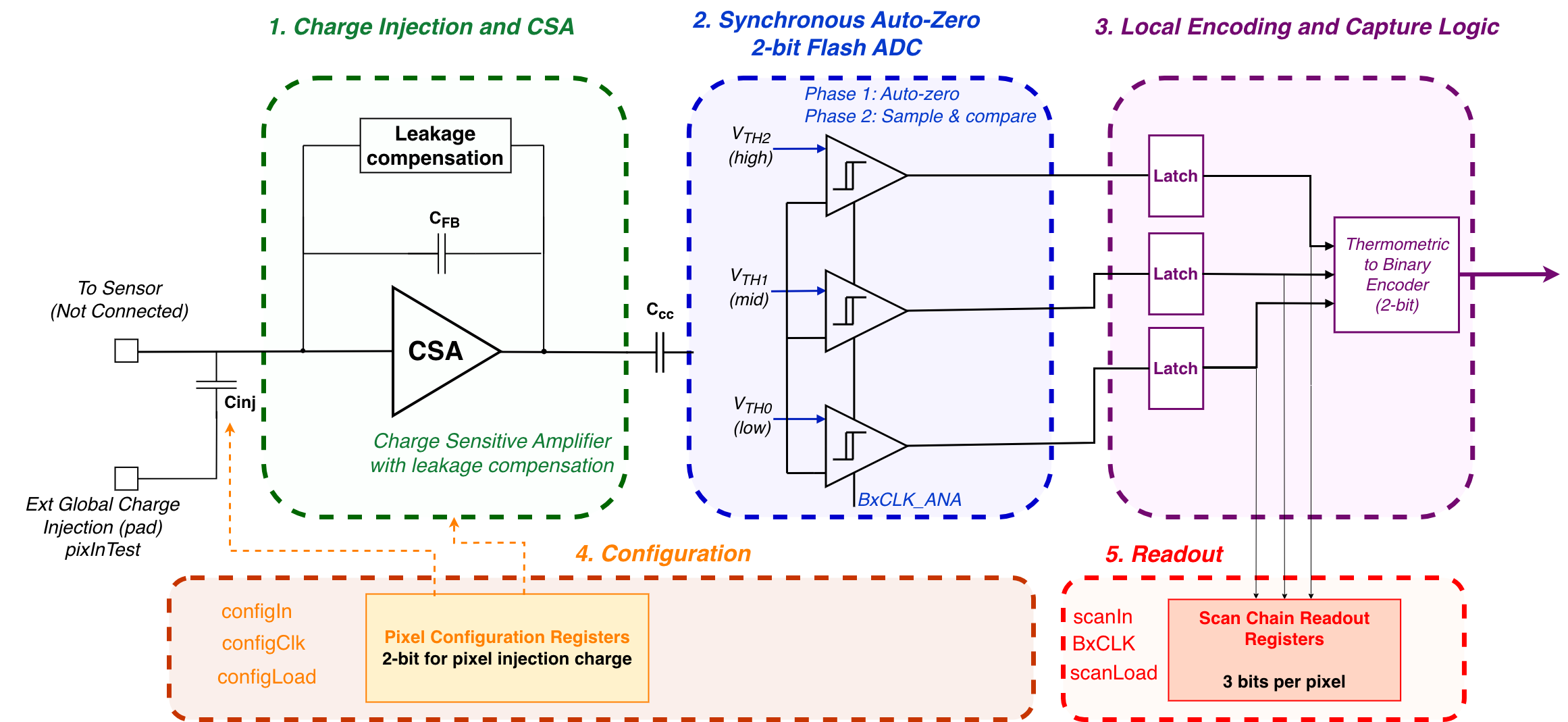}
        \caption{Pixel-level signal-processing architecture.}
        \label{fig:ROICpix_a}
    \end{subfigure}

    \vspace{0.8em}

    \begin{subfigure}[t]{0.98\textwidth}
        \centering
        \includegraphics[
            width=\linewidth,
            height=0.3\textheight,
            keepaspectratio
        ]{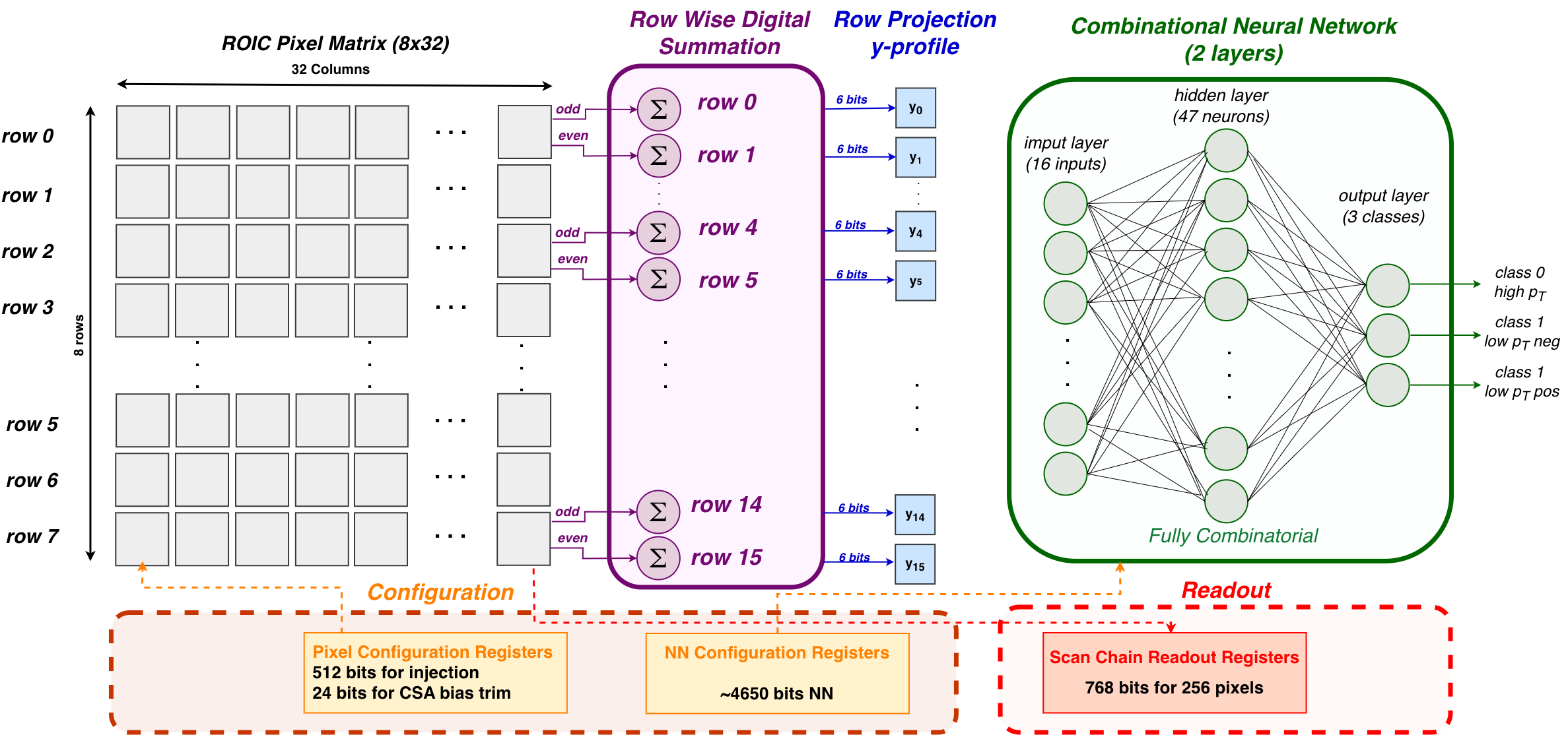}
        \caption{Superpixel-level feature formation and neural network inference.}
        \label{fig:ROICpix_b}
    \end{subfigure}

    \vspace{0.8em}

    \begin{subfigure}[t]{0.98\textwidth}
        \centering
        \includegraphics[
            width=\linewidth,
            height=0.235\textheight,
            keepaspectratio
        ]{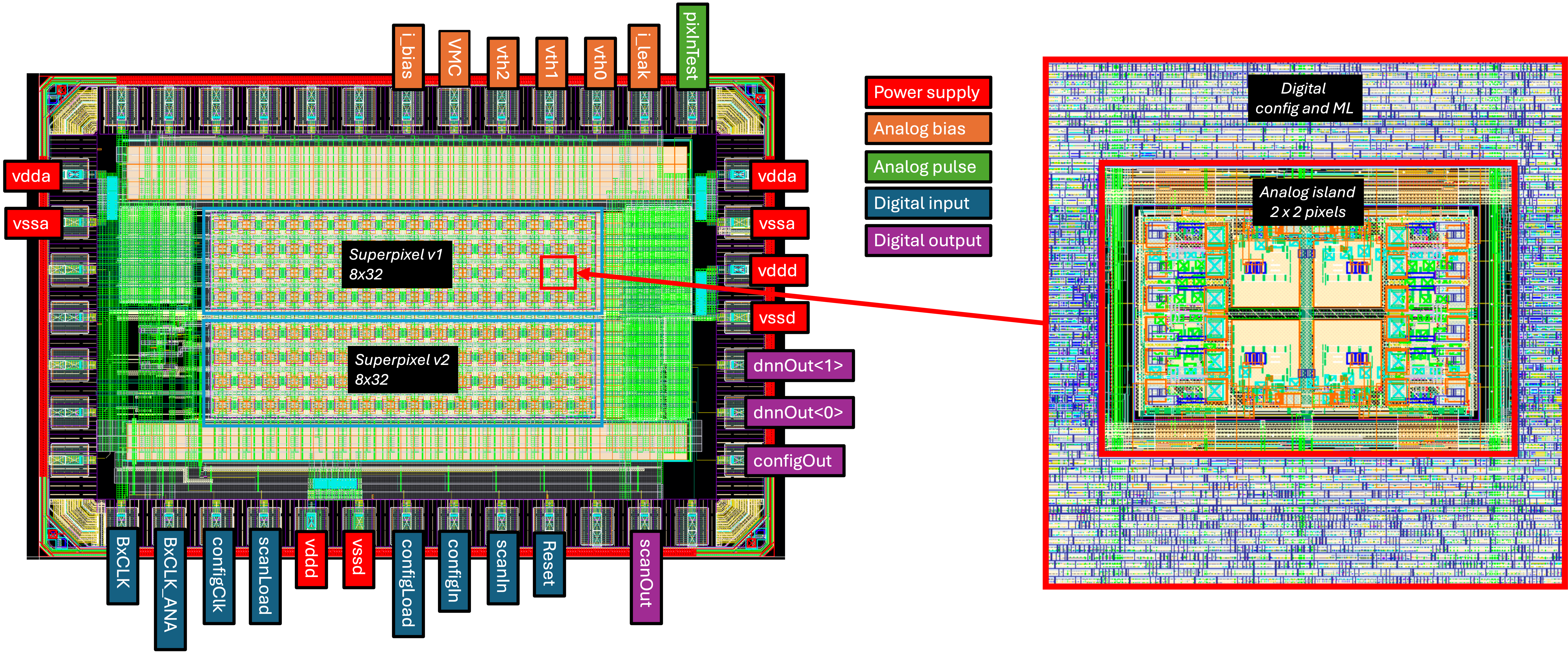}
        \caption{Full-chip physical implementation (left) with an enlarged view of the
        $2\times2$ pixel block (right).}
        \label{fig:ROICpix_c}
    \end{subfigure}

    \caption{\smartpixels architecture and physical implementation.
(a) Pixel topology comprising the charge-sensitive amplifier, synchronous auto-zero 2-bit flash ADC, thermometric-to-binary encoder, configuration logic, and scan-chain registers.
(b) Superpixel topology, in which the encoded pixel outputs are summed into 16 six-bit row projections and processed by a fully combinational neural network.
(c) Full-chip layout showing the two $8\times32$ superpixels and an enlarged $2\times2$ pixel block with its central analog island and surrounding digital logic. The prototype contains 256 pixels per superpixel and 512 pixels in
total.}
    \label{fig:ROICpix}

    \vspace*{\fill}
\end{figure*}

A prototype readout integrated circuit (ROIC) was implemented as a $1 \times 1.6$\,mm$^2$ ASIC in a 28\,nm TSMC CMOS process. The design comprises two $32 \times 8$ arrays of $25 \times 25$\,\textmu m$^2$ pixels, referred to as \emph{superpixels} (SP). Each pixel integrates an analog front-end (AFE) for signal processing together with digital back-end logic for cluster classification. The pixel- and superpixel-level architectures, together with their physical implementation, are shown in Fig.~\ref{fig:ROICpix}. The AFE consists of a charge-sensitive preamplifier that integrates the sensor signal, followed by a 2-bit flash analog-to-digital converter (ADC) which samples the integrated signal at the bunch-crossing clock (BxCLK) and produces three thermometrically encoded comparator outputs.

The use of bunch-synchronous digitization is a central feature of the proposed architecture. In contrast to conventional pixel readout schemes that estimate the deposited charge from the discriminator time over threshold (ToT), all pixels are sampled using a common temporal reference. The digitized values from neighboring pixels are therefore associated with the same bunch crossing, preserving the spatial charge distribution and its temporal correlations before the data are passed to the downstream cluster-processing logic.

This approach provides several system-level advantages ~\cite{ISCAS2023,Parpillon:2024TWEPP,parpillon2024smartpixelsinpixelai}. First, it preserves the temporal structure of the particle cluster, which is required by the back-end digital algorithm.  Second, the ADC produces a fixed-rate and deterministic representation of the pixel response, avoiding the need to reconstruct charge from discriminator-edge timing and pulse duration. Third, provided that the Charge Sensitive Amplifier (CSA) remains within its dynamic range and settles sufficiently between samples, signals from consecutive bunch crossings can be digitized independently with reduced ambiguity from overlapping discriminator pulses. The resulting bunch-resolved data format is naturally suited to low-latency combinational processing and to real-time physics-driven data selection directly within the detector readout.

The synchronous architecture does not eliminate pile-up arising from CSA saturation or incomplete signal recovery. It does, however, avoid the additional ambiguity introduced when charge is encoded through the duration of potentially overlapping ToT pulses. Realizing this bunch-resolved representation locally within each pixel also requires the digitization circuitry to satisfy stringent area and power constraints. The ADC resolution therefore represents a tradeoff between the amount of charge information retained for cluster classification and the complexity of the analog implementation. Previous work~\cite{Yoo_2024} demonstrated that, for the cluster-classification task considered here and the adopted pixel pitch, a 2-bit ADC provides sufficient information to achieve 54--75\% data reduction while enabling a compact, low-complexity analog implementation, and 3 bits increases the area substantially with only minimal improvement in performance.
Requiring only three comparators, the 2-bit flash ADC allows the complete digitization stage to be integrated within the $25\times25~\mu\mathrm{m}^{2}$ pixel footprint while remaining compatible
with the available area and power budgets. The physical implementation is shown in Fig.~\ref{fig:ROICpix_c}, including the full-chip layout and an enlarged view of a $2\times2$ pixel block that highlights the central analog island and the surrounding digital logic.

The three comparator thresholds are supplied off-chip via the bias voltages V$_{\mathrm{th0}}$, V$_{\mathrm{th1}}$, and V$_{\mathrm{th2}}$, which control the tripping points of bits 0, 1, and 2, respectively. The ADC operates in two successive phases: an auto-zero phase, followed by sampling of the integrated charge. 

As shown in the full-chip layout of Fig.~\ref{fig:ROICpix_c}, two pixel variants were fabricated to experimentally compare alternative ADC architectures: superpixel~v1 (SP1) employs a differential comparator architecture,  and superpixel~v2 (SP2) uses a single-ended implementation. The differential architecture was expected to provide improved immunity to common-mode disturbances and supply variations, while the single-ended architecture was designed to minimize pixel area and analog power. Their measured performance is compared in Section IV.

Due to the limited ASIC area, the prototype was not bump bonded to a sensor. Instead, a dedicated test input, \texttt{pixInTest}, is provided to emulate charge generation within the ASIC.

\subsection{Digital Feature Formation and Neural Network Inference}
The superpixel-level feature formation and classification datapath is
illustrated in Fig.~\ref{fig:ROICpix_b}. During normal operation, the digital data from each pixel is latched and encoded into a 2-bit binary format. The outputs are digitally summed along columns while preserving row information, thereby reducing the 2-bit data from 256 pixels into 16 buses of 6-bit values.  
This effectively projects the raw pixel data along the local $y$-axis. The resulting 16 buses are subsequently processed by the digital logic, which implements the neural network classifier.

The entire back-end digital chain, from encoding through classification, is realized using combinational logic which offers two major advantages. First, the absence of sequential logic minimizes inference latency, allowing classification to be completed within a single \texttt{BxCLK} period. Second, dynamic power is consumed only when the input pattern changes. Since the expected pixel occupancy in HL-LHC tracking detectors is below 0.2\%, switching activity is sparse, substantially reducing the average digital power consumption.

\subsection{Configuration and Test Readout}
The ASIC includes dedicated configuration and scan-chain interfaces for programming the charge-injection and internal bias settings and for characterizing the analog and digital signal paths. The pixel-level configuration and scan-chain paths are shown in
Fig.~\ref{fig:ROICpix_a} and Fig.~\ref{fig:ROICpix_b}. Configuration data are loaded serially, while the thermometric comparator outputs can be captured independently during test by the scan-chain registers and shifted off chip using the external \texttt{BxCLK}.

The back-end digital logic includes a compact two-layer NN that classifies charge clusters based on the $p_{\mathrm{T}}$ of the incident charged particles.
The model architecture, training procedure, and dataset generation are described in~\cite{Yoo_2024, shekar2025}. The NN was trained on charge profiles from simulated pion tracks with kinematic properties taken from CMS~\cite{CMS_2008} collision data~\cite{zenodo_16x16}. 

To enable an efficient hardware implementation, the network was quantized using \texttt{QKeras} and translated into synthesizable RTL using the \texttt{hls4ml} framework and Siemens Catapult HLS before integration into the digital back-end.

The following sections experimentally characterize the AFE and validate the complete mixed-signal signal chain in silicon.

\section{DAQ and Testing Setup}
\label{sec:setup}

\begin{figure}[t!]
    \centering
    \includegraphics[width=0.5\linewidth]{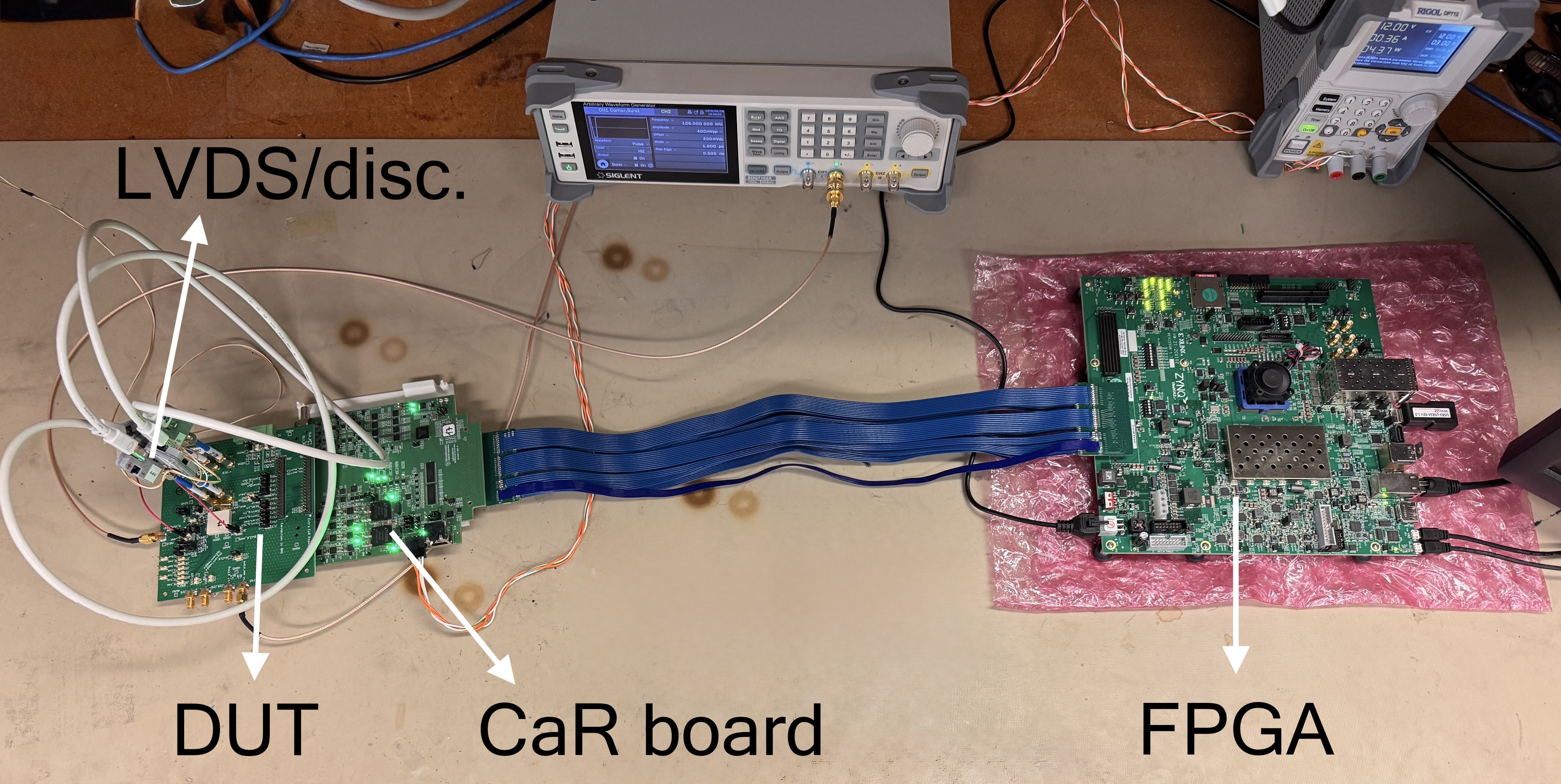}
    \caption{The device under test (DUT), a custom PCB with the bonded ROIC, connects via a \textsc{SEARAY} connector to the \textsc{CaR} board, which interfaces with a Xilinx ZCU102 System on Chip FPGA for communication to a workstation running the Python test routines (not shown). External power supplies bias the DUT and \textsc{CaR} board. An external generator supplies precise pulses.}
    \label{fig:testStand}
\end{figure}

A data acquisition (DAQ) system was developed to characterize the ASIC using the open-source \textsc{Caribou}~\cite{Vanat:2703500} and \textsc{Spacely}~\cite{Quinn:2024xhl} frameworks. The system is shown in Fig.~\ref{fig:testStand}. A host PC runs the \textsc{Spacely} software, which executes test routines written in Python that communicate with a Xilinx ZCU102 System-on-Chip (SoC) FPGA. Custom FPGA firmware generates control waveforms to operate the ASIC, which are routed through a Control and Readout (\textsc{CaR}) board~\cite{Vanat:2703500}, a custom PCB designed by the \textsc{Caribou} Project. The \textsc{CaR} board converts FPGA signals into clean digital control signals and provides power, bias voltages, and analog inputs to the device under test (DUT). 

While the \textsc{CaR} board can generate analog pulses, higher-quality stimuli were supplied using a 5~GSa/s, 14-bit SDG7102A waveform generator. The DUT board, designed at Fermilab, hosts the ASIC and enables bi-directional communication with the DAQ system. This setup has been successfully replicated at multiple testing sites: Fermilab, The University of Chicago, and Cornell University.

The \textsc{CaR} board was selected as a lightweight and relatively low-cost test platform that can be readily replicated and shared across collaborating institutions. Its digital signal path, however, has a usable bandwidth of approximately 12~MHz, limiting the present measurements to a 10~MHz \texttt{BxCLK} rather than the desired operational frequency of 40~MHz.

\section{Analog Front-End Characterization}
\label{sec:afe_results}

The objective of this section is to establish the operating point, quantify the analog performance of the prototype, and identify the dominant limitations revealed by first-silicon measurements.

The characterization was planned in four different stages. First, the optimum operating point is established by minimizing the equivalent noise charge (ENC) as a function of bias current. The front-end transfer characteristic is then measured to determine its conversion gain and usable dynamic range. With the operating point fixed, the intrinsic noise and threshold uniformity are quantified across the pixel matrix. Finally, the dominant non-idealities observed in first-silicon are analyzed to guide the design of the next-generation prototype.

\subsection{Procedure}

The AFE was characterized using calibrated charge injection through the programmable on-chip injection capacitor. Charge was injected synchronously with the \texttt{BxCLK}, and the resulting comparator outputs were sampled and recorded through the scan-chain readout. Unless otherwise noted, all measurements were performed at room temperature at a \texttt{BxCLK} frequency of 10~MHz (owing to limitations of the \texttt{CaR} board as discussed in Sec.~\ref{sec:setup}). The 10~MHz operating point does not limit our ability to characterize the front-end noise, conversion gain, threshold dispersion, and pixel-to-pixel uniformity, because these quantities are expected to depend only weakly on the \texttt{BxCLK} frequency, provided that the front end settles before sampling. Operation at 40~MHz may require a modest increase in bias current to satisfy the shorter settling-time requirement, but is not expected to substantially alter the analog performance metrics reported here. The 10~MHz measurements therefore provide a representative characterization of the front end, while measurements at the nominal frequency remain necessary to validate timing and establish the final operating bias current.

S-curve measurements constitute the primary characterization technique used throughout this work because they simultaneously provide threshold and noise information for every comparator in the pixel matrix. For each pixel and each ADC comparator, the probability of a comparator transition was measured as a function of injected charge. The resulting transition was fitted with a Gaussian cumulative distribution function, from which the mean ($\mu$) and standard deviation ($\sigma$) were extracted. Across the pixel matrix, the distribution of $\mu$ values represents the threshold dispersion, while the distribution of $\sigma$ values corresponds to the equivalent noise charge (ENC).

\begin{figure}[t!]
    \centering
    \includegraphics[width=0.5\linewidth]{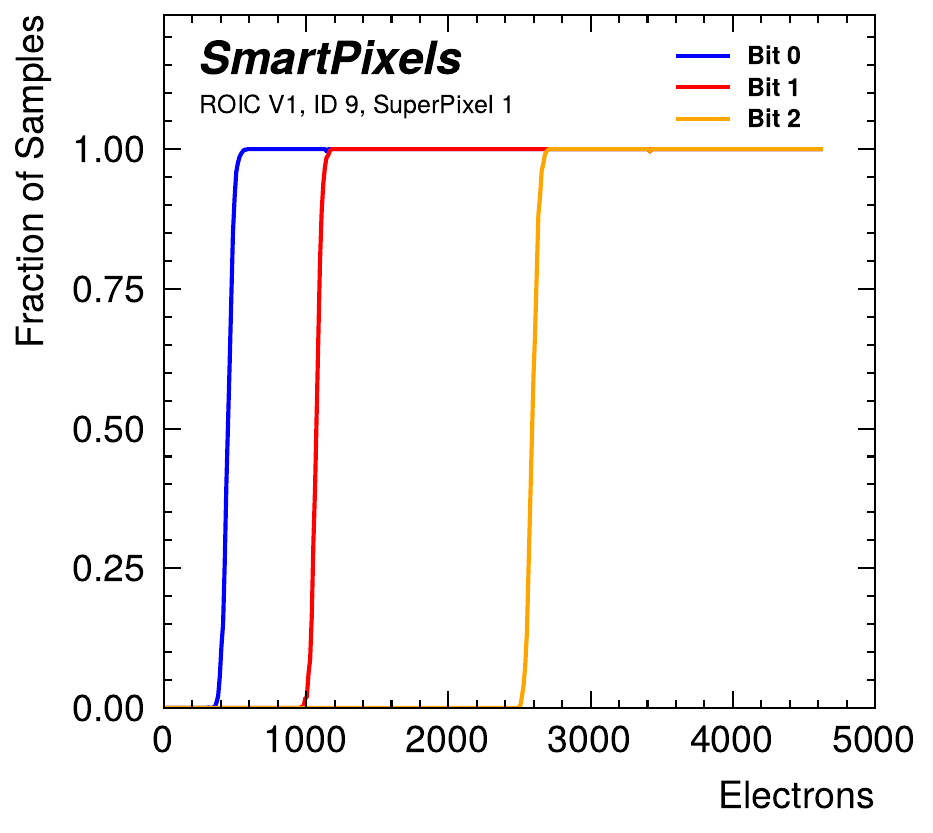}
    \caption{Representative S-curve measured for a single pixel. The probability of comparator firing is plotted as a function of injected charge. The Gaussian cumulative distribution fit is used to extract the threshold ($\mu$) and equivalent noise charge (ENC, $\sigma$).}
    \label{fig:Scurve}
\end{figure}

The AFE was characterized primarily through repeated S-curve measurements. A typical S-curve plot is shown in Fig.~\ref{fig:Scurve}.

The injected charge was calculated from the pulse amplitude according to
\begin{equation}
Q_{\mathrm{inj}}=\frac{V_{\mathrm{inj}}\,C_{\mathrm{inj}}}{q},
\end{equation}

where $V_{\mathrm{inj}}$ is the pulse amplitude, $C_{\mathrm{inj}}$ is the programmable on-chip injection capacitance, and $q = 1.602\times10^{-19}$~C is the elementary charge. Unless otherwise noted, the injection capacitance was configured to its minimum value of 1.85~fF.

\subsection{Operating point (bias current)}

The analog bias current establishes the operating point of the front end and is a primary design parameter governing gain, bandwidth, noise, and power consumption. Its nominal range was selected during the design phase to satisfy the stringent power constraints of HL-LHC pixel detectors~\cite{RD53Bspec}. A total power density of approximately 1~W/cm$^2$ is typically allocated to the innermost tracking layers, including both analog and digital circuitry. For the adopted pixel pitch of $25\times25~\mu\mathrm{m}^2$, this corresponds to a total power budget of approximately 6.25~$\mu$W per pixel. During the design phase, approximately 5~$\mu$W per pixel was allocated to the AFE and 1~$\mu$W per pixel to the digital processing chain. Within this power envelope, circuit simulations predicted a favorable operating range of 3--5.5~$\mu$A per pixel. Because all subsequent analog and digital measurements depend on the front-end operating conditions, the bias current was optimized before the remaining characterization was performed. For the measurements reported here, the objective was to minimize the equivalent noise charge (ENC), thereby providing the highest analog signal quality for evaluating the complete mixed-signal processing chain.

\begin{figure}[t!]
\centering
\includegraphics[width=0.5\linewidth]{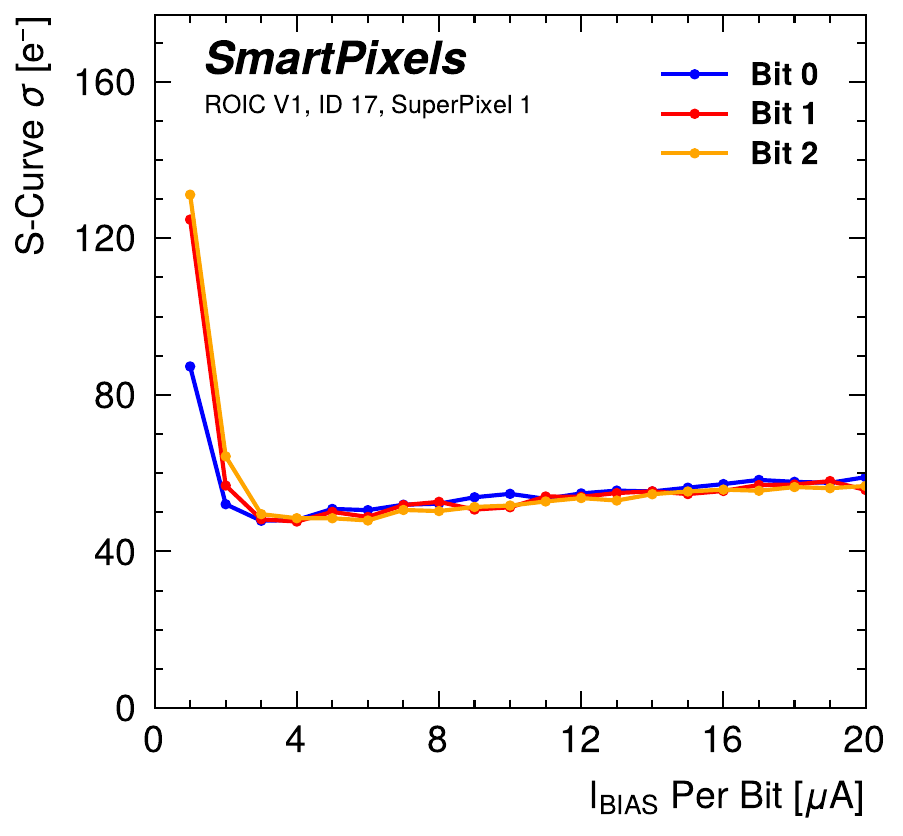}
\caption{Measured ENC as a function of SP1 pixel bias current. The minimum ENC occurs at a bias current of approximately 3.15~$\mu$A per pixel.}
\label{fig:Ibias}
\end{figure}

As shown in Fig.~\ref{fig:Ibias}, bias currents below approximately 2~$\mu$A per pixel result in a rapid degradation of noise performance. In this regime, insufficient bias reduces the preamplifier open-loop gain and increases the input-referred thermal-noise contribution. The minimum measured ENC is approximately 47.6~\unit{\electron}, obtained at bias currents of 3.15~$\mu$A per pixel for SP1 and 2.2~$\mu$A per pixel for SP2. At higher currents, the ENC increases gradually, reaching approximately 59~\unit{\electron} at 20~$\mu$A per pixel. Thus, increasing the bias beyond the measured optimum does not provide an additional noise-performance benefit under the tested conditions. The measured optimum occurs near the lower end of the simulated operating range. This shift is likely related to the reduced bunch-crossing frequency of 10~MHz used during characterization. At this frequency, the settling-time and bandwidth requirements are less stringent than at the nominal 40~MHz operating frequency, allowing the front-end to operate at a lower bias current. A modest increase in bias may therefore be required at 40~MHz to maintain the required settling accuracy. Based on the measured dependence on bias current, such an adjustment is expected to have only a limited effect on the ENC. The conversion gain and threshold dispersion are also expected to remain largely unchanged, since they primarily depend on the charge gain, device mismatch, and bias-distribution uniformity rather than on the sampling frequency. Measurements at 40~MHz are nevertheless needed to confirm these expectations and establish the final detector operating point. A bias current of 3.15~$\mu$A per pixel was therefore selected for SP1 throughout the remainder of the study. This operating point provides the lowest measured ENC and establishes favorable analog conditions for evaluating the functionality of the complete mixed-signal architecture; it should not be interpreted as the optimized operating point for nominal 40~MHz operation. At the selected bias current, the AFE dissipates approximately 0.45~W/cm$^2$, corresponding to 2.8~$\mu$W per pixel. This value is well below the allocated analog budget of approximately 5~$\mu$W per pixel and leaves substantial margin within the total 6.25~$\mu$W-per-pixel system budget for the digital processing circuitry.

\subsection{Conversion Gain}

The conversion gain (CvG) defines the relationship between collected charge and comparator threshold voltage and is a key parameter for calibrating the front end, expressing the measured noise in electrons, and evaluating pixel-to-pixel response uniformity.

\begin{figure*}[t!]
    \centering
    \begin{subfigure}[b]{0.48\linewidth}
        \centering
        \includegraphics[width=\linewidth]{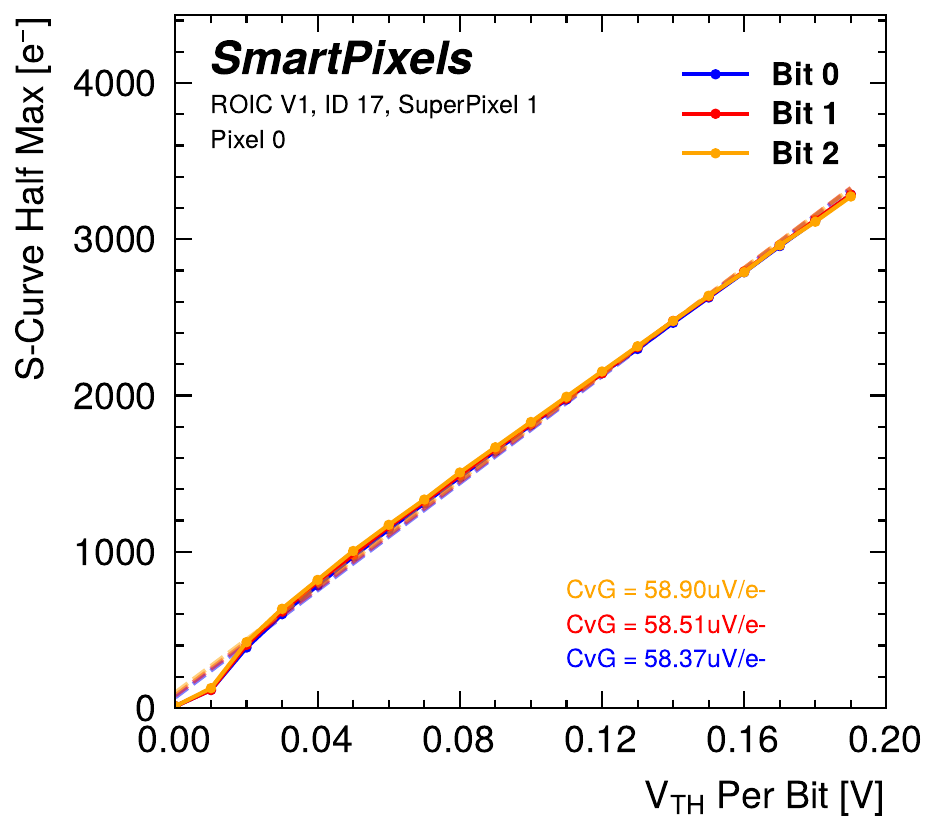}
        \caption{Low and mid charge injected region for SP1}
        \label{fig:ID17SP1Linearity}
    \end{subfigure}
    \hfill
    \begin{subfigure}[b]{0.48\linewidth}
        \centering
        \includegraphics[width=\linewidth]{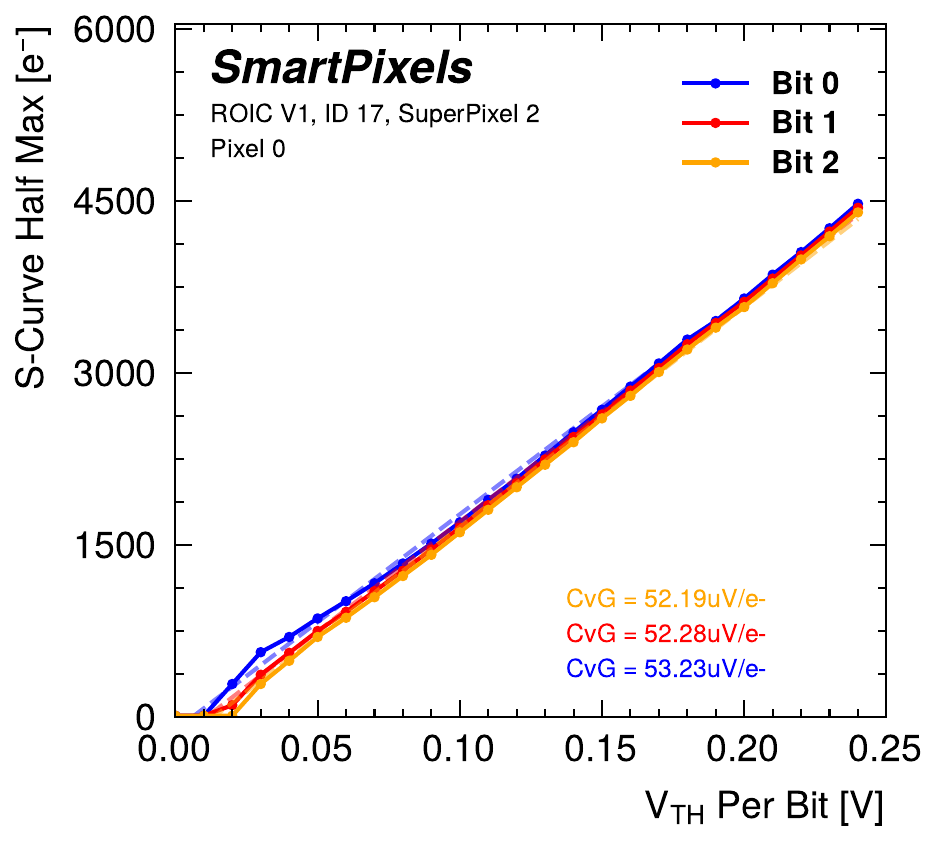}
        \caption{Low and mid charge injected region for SP2}
        \label{fig:ID17SP2Linearity}
    \end{subfigure}
    
    \begin{subfigure}[b]{0.48\linewidth}
        \centering
        \includegraphics[width=\linewidth]{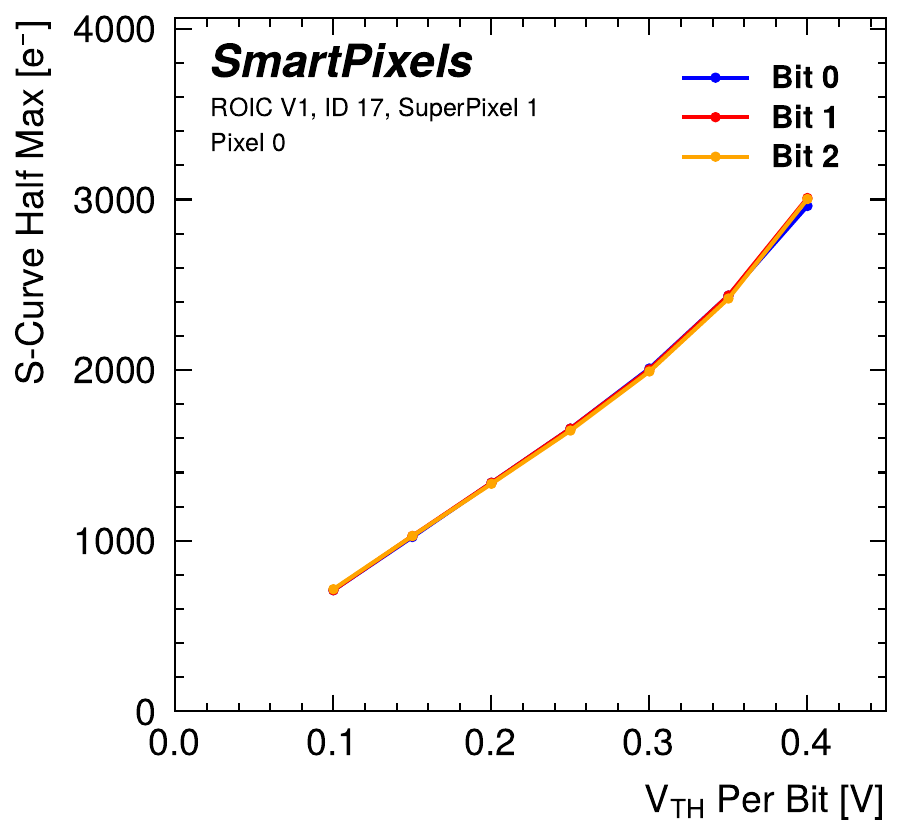}
        \caption{High charge injected region for SP1}
        \label{fig:ID17SP1LinearityHigh}
    \end{subfigure}
    \hfill
    \begin{subfigure}[b]{0.48\linewidth}
        \centering
        \includegraphics[width=\linewidth]{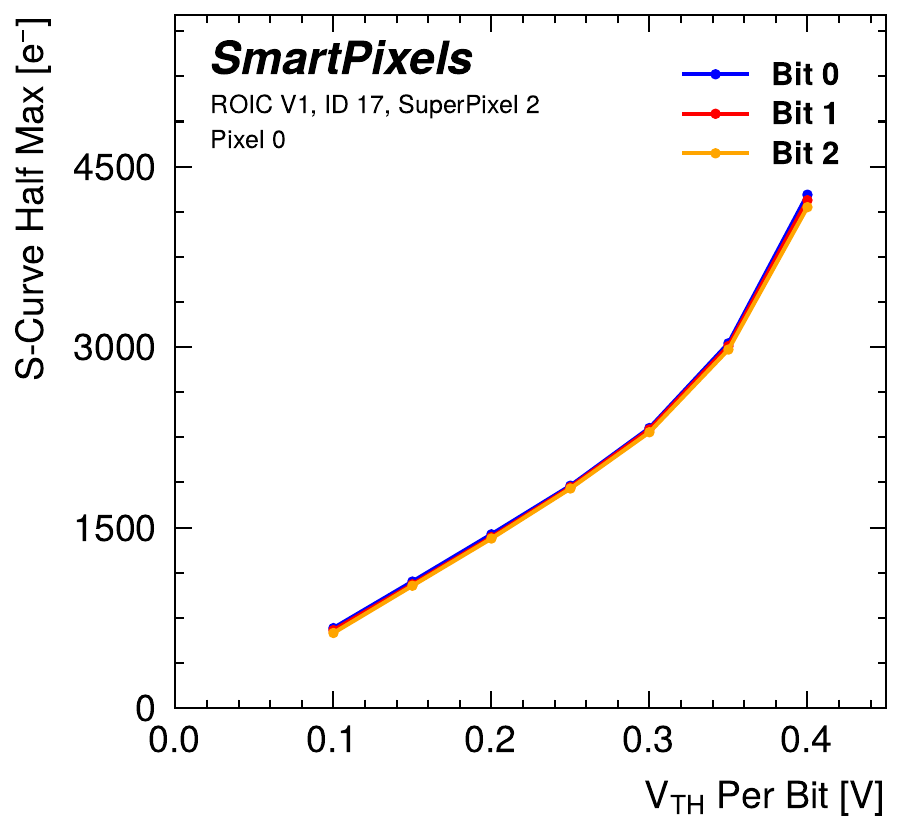}
        \caption{High charge injected region for SP2}
        \label{fig:ID17SP2LinearityHigh}
    \end{subfigure}
    
    \caption{Measured transfer characteristic of the AFE for the SP1 and SP2 architectures. The S-curve mean ($\mu$) is plotted as a function of the applied comparator threshold voltage for a representative pixel. Three operating regimes are observed: (i) low-threshold nonlinearity caused by comparator sampling-switch charge injection, (ii) a linear operating region used to extract the conversion gain, and (iii) high-charge saturation due to preamplifier compression. The dashed lines indicate the linear fits used to determine the CvG.}
    \label{fig:LinearityComparison}
\end{figure*}

The CvG was measured by extracting the S-curve mean as a function of the applied comparator threshold voltage and fitting the linear portion of the response. The measured transfer characteristics are shown in Fig.~\ref{fig:LinearityComparison}, which illustrates the relationship between the applied comparator threshold voltage and the injected charge used to extract the conversion gain (CvG).

Three operating regimes can be identified. At thresholds $<$400~\unit{\electron}, the response exhibits significant nonlinearity due to charge injection from the comparator sampling switch. During the sampling phase, closure of this switch deposits charge on the sampling capacitor, producing an effective threshold offset of approximately 3~mV. This contribution is most significant at low signal amplitudes and becomes progressively less important as the applied threshold increases. 

Over the operating range relevant for detector applications, the response is approximately linear from 500 to 8000~\unit{\electron} for SP1 and from 800 to 8000~\unit{\electron} for SP2. The slope measured in this region defines the conversion gain. 

Above approximately 8000~\unit{\electron}, the response departs from linearity due to saturation of the preamplifier, consistent with circuit simulations. 

\begin{figure}[t!]  
    \centering
    \includegraphics[width=0.5\linewidth]{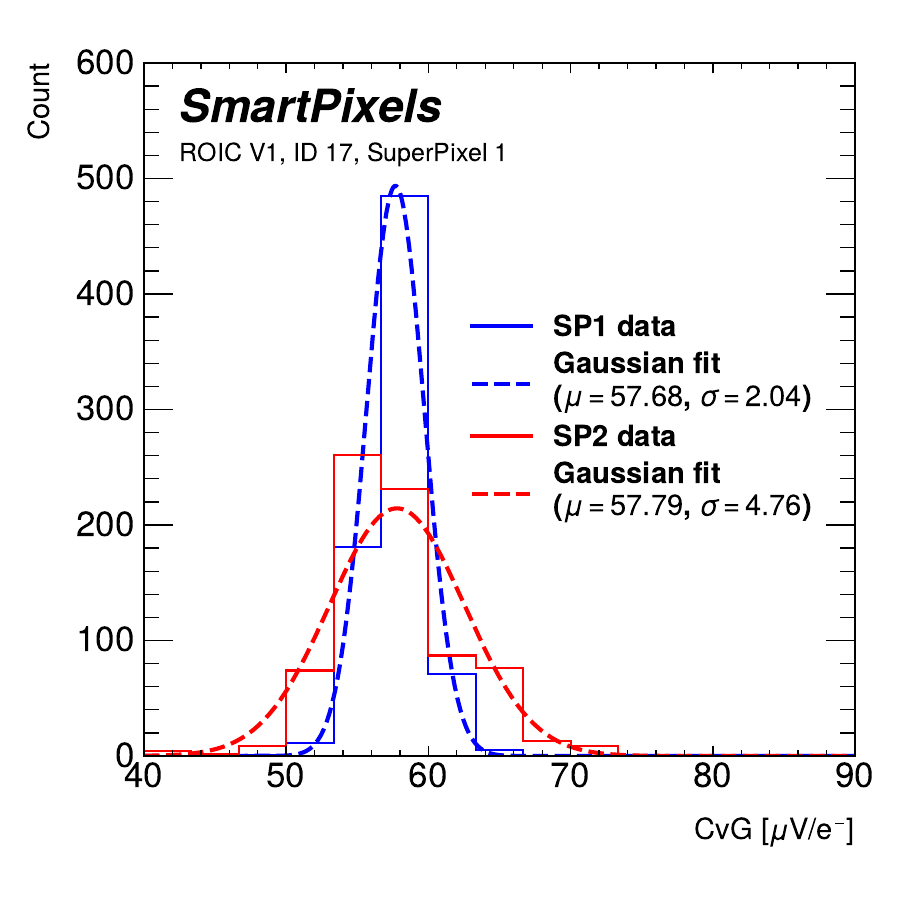}
    \caption{
Distribution of the extracted conversion gain across all 256 pixels for the SP1 and SP2 architectures. Gaussian fits are overlaid to determine the mean conversion gain and pixel-to-pixel dispersion.}
    \label{fig:CombinedCvG}
\end{figure}

The CvG was extracted for every pixel in both superpixel variants. The resulting distributions are shown in Fig.~\ref{fig:CombinedCvG}.
The conversion gain was measured across the full pixel matrix to quantify pixel-to-pixel variations. The mean CvG is 57.68~\unit{\micro\volt\per\electron} for SP1 (differential) and 57.79~\unit{\micro\volt\per\electron} for SP2 (single-ended), with standard deviations of 2.04 and 4.76~\unit{\micro\volt\per\electron}, respectively. Although the two architectures exhibit comparable mean conversion gains, SP2  shows approximately twice the pixel-to-pixel dispersion of SP1. A larger dispersion is consistent with the reduced common-mode rejection of the single-ended architecture which inherently leads to greater sensitivity to local device and bias variations. However, the magnitude of the observed difference was not predicted, and subsequent measurements identified a likely implementation-related contribution associated with the SP2 threshold bias path, which is discussed in Sec.~\ref{sec:qth_enc}. A larger CvG dispersion increases pixel-to-pixel threshold variation after calibration and therefore reduces the uniformity of the digitized detector response. The measured variations remain sufficiently small to support charge discrimination and the extraction of ENC and threshold dispersion over the operating range used in this work.

\subsection{Threshold Dispersion and ENC}\label{sec:qth_enc}

The performance of the intelligent pixel architecture ultimately depends on the stability of the AFE. The proposed architecture relies on multiple discriminator thresholds to generate a multi-level representation of the collected charge that serves as the input to the digital classifier. Consequently, the equivalent noise charge (ENC), threshold dispersion, and baseline stability directly determine the fidelity of the digitized charge information presented to the downstream digital processing.

Early simulations in~\cite{Yoo_2024} indicated that the classifier achieved a desirable efficiency when the lowest discriminator threshold (V$_{\mathrm{th0}}$) was set to approximately 400~\unit{\electron}, to maximize the signal collected while maintaining acceptable noise occupancy. Operation at substantially higher thresholds progressively removes useful cluster information and degrades classification performance.

To ensure reliable operation, the lowest discriminator threshold should remain approximately four standard deviations above the total rms charge uncertainty which is defined as

\begin{equation}
\sigma_Q=
\sqrt{\mathrm{ENC}^2+
\sigma_{Q_{\mathrm{th}}}^2+
\sigma_{\mathrm{BL}}^2},
\label{eq:sigmaQ}
\end{equation}

where $\sigma_{Q_{\mathrm{th}}}$ denotes the threshold dispersion and
$\sigma_{\mathrm{BL}}$ represents baseline fluctuations which is a negligible contribution in our design.

\begin{equation}
Q_{\mathrm{th,min}}
\gtrsim
4\sigma_Q ,
\label{eq:threshold}
\end{equation}

Assuming Gaussian-distributed fluctuations, a discriminator threshold located at $4\sigma_Q$ corresponds to a theoretical noise occupancy of approximately
$3.2\times10^{-5}$ false hits per discriminator decision.

To satisfy this requirement, the AFE was designed to operate with a minimum discriminator threshold of approximately 400~\unit{\electron}. Circuit simulations predict an equivalent noise charge (ENC) of approximately 40~\unit{\electron} for the intrinsic front end and approximately 55~\unit{\electron} under the measurement conditions used in this work. When connected to the target detector capacitance of 30~fF, the ENC increases to approximately 88~\unit{\electron}, consistent with the expected capacitive noise scaling. The design further targets a threshold dispersion of approximately 45~\unit{\electron} at room temperature.

\begin{equation}
\sigma_Q
=
\sqrt{88^2+45^2}
\approx
100~\unit{\electron},
\end{equation}

For a minimum operating threshold of approximately
400~\unit{\electron}, this corresponds to a design target of
$$
\sigma_Q \lesssim 100~\unit{\electron}.
$$

The prototype characterized in this work was not bump bonded to a sensor and therefore measures the intrinsic front-end performance. Consequently, agreement between the measured and simulated ENC directly validates the analog noise performance of the fabricated charge-sensitive amplifier, while discrepancies in threshold dispersion primarily reflect first-silicon implementation effects rather than limitations of the AFE architecture.

\begin{figure*}[t!]
    \centering

    \begin{subfigure}[t]{0.47\textwidth}
        \centering
        \includegraphics[width=\linewidth]{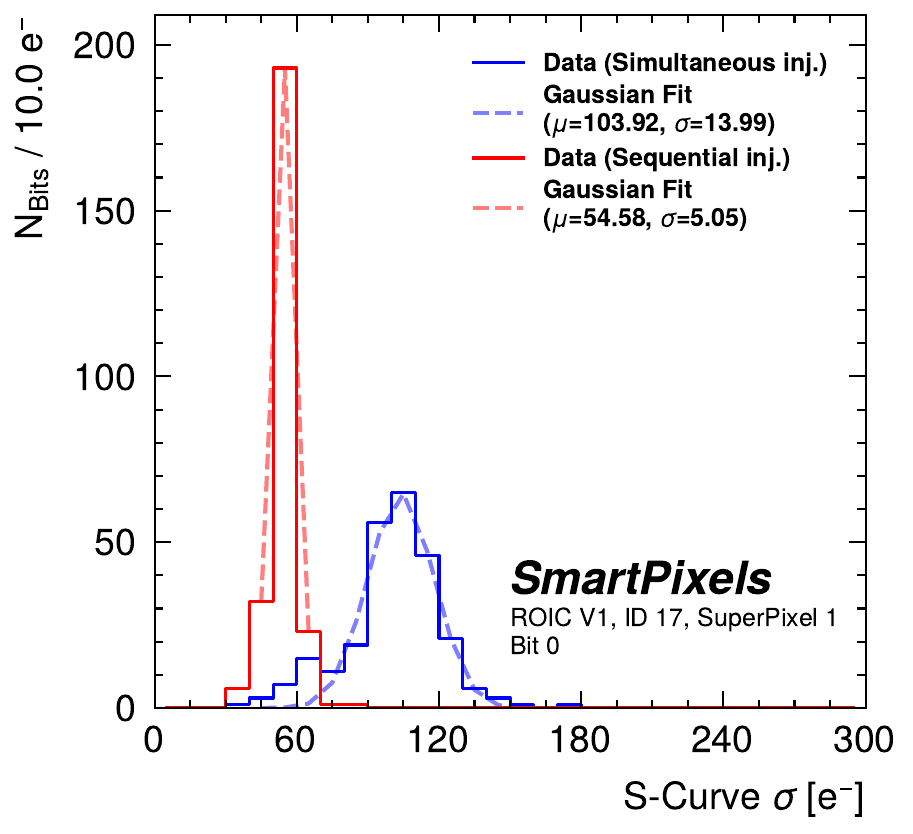}
        \caption{ENC distribution with all pixels injected simultaneously (``simultaneous inj.") and sequentially (``sequential inj.").}
        \label{fig:enc_all_pixels}
    \end{subfigure}
    \hfill
    \begin{subfigure}[t]{0.47\textwidth}
        \centering
        \includegraphics[width=\linewidth]{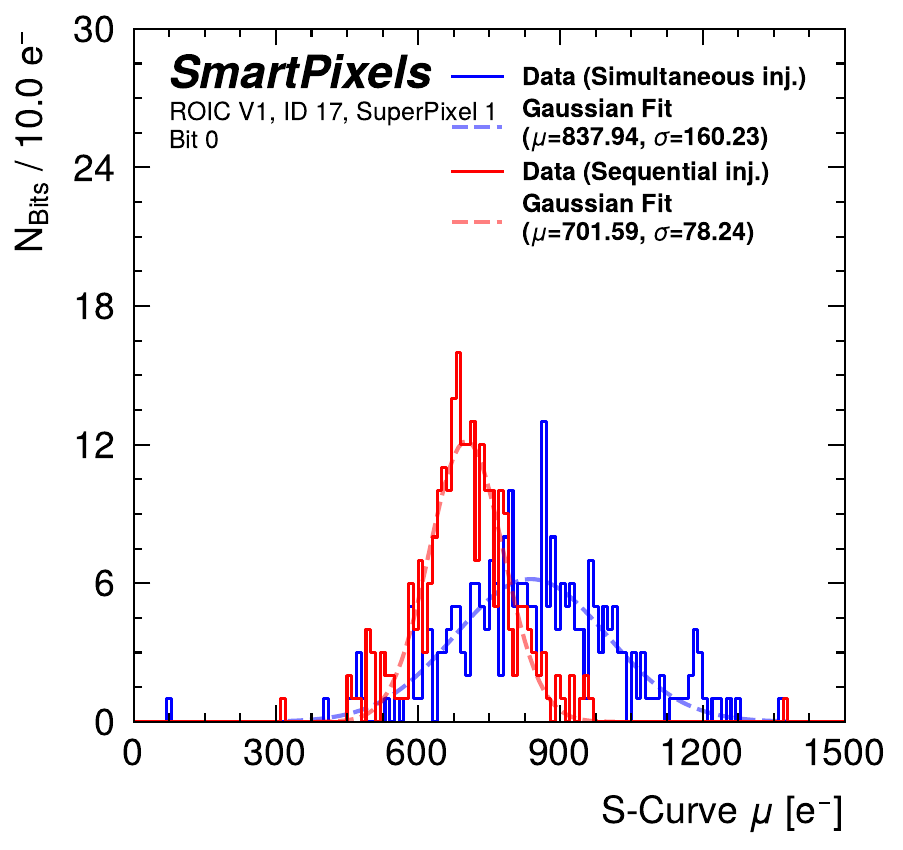}
        \caption{Threshold distribution with all pixels injected simultaneously (``simultaneous inj.") and sequentially (``sequential inj.").}
        \label{fig:qth_all_pixels}
    \end{subfigure}

    \caption{Measured ENC and threshold distributions for SP1 under nominal
    operating conditions. Blue gaussian plots show measurements obtained with all
    pixels injected simultaneously. The aggregate capacitive loading of the
    shared charge-injection network distorts the calibration waveform,
    increasing the apparent ENC from approximately 54.6~\unit{\electron} to
    104~\unit{\electron} and the threshold dispersion from approximately
    78.2~\unit{\electron} to 160.2~\unit{\electron}. The red Gaussian-fit curves show sequential
    measurements obtained with one injection site active at a time. This procedure substantially reduces, but does not completely eliminate, pulse-loading effects and therefore provides a less distorted estimate of the
    pixel response. 
    All measurements were performed with
    $V_{\mathrm{th}}=0.031$~V, corresponding to an effective threshold of
    approximately 700~\unit{\electron}.}
    \label{fig:enc_and_thresh}
\end{figure*}

The measured ENC and threshold-dispersion distributions are presented in
Fig.~\ref{fig:enc_and_thresh}. Two charge-injection configurations are
compared to distinguish the intrinsic front-end response from artifacts
introduced by the global calibration network.

When all pixels are pulsed simultaneously, the global calibration line drives
the aggregate injection load of the full matrix in addition to the fixed
parasitics of the distribution network. The resulting distributed RC loading
distorts the local pulse waveform and causes the effective injected charge to
differ from that inferred from the nominal calibration settings. This effect
depends on the aggregate switched load, the distributed network impedance,
and the local injection path. Nonuniform effective charge directly broadens
the distribution of fitted S-curve means. In addition, pulse-shape and timing
variations of the loaded calibration network can broaden the individual
S-curves and therefore appear as an increased ENC. With simultaneous
all-pixel injection, the apparent ENC and threshold dispersion are
approximately 104~\unit{\electron} and 160~\unit{\electron}, respectively.

To reduce this calibration-induced contribution, the pixels were subsequently
characterized sequentially, with only one injection site active and the
calibration setting reduced to its minimum value. This substantially reduced
the actively switched load, although the fixed parasitics of the global
distribution network remain. Under these conditions, the measured average
ENC decreases to 54.6~\unit{\electron} and the threshold dispersion decreases to 78.24~\unit{\electron}. The measured ENC agrees closely with the simulated value of approximately 55~\unit{\electron} for the front end without an attached sensor,
supporting the simulated intrinsic noise performance of the fabricated CSA.
For comparison, circuit simulations predict an ENC of approximately
88~\unit{\electron} when the front end is loaded by the target detector capacitance
of 30~fF.

The threshold dispersion measured with sequential injection remains larger
than the simulated mismatch-limited value of approximately 45~\unit{\electron}.
Assuming statistically independent contributions, the difference corresponds
to an additional effective dispersion of approximately

\begin{equation}
\sigma_{\mathrm{add}}
=
\sqrt{(78.24~\unit{\electron})^2-(45~\unit{\electron})^2}
\approx 63.9~\unit{\electron}.
\end{equation}

Several first-prototype effects may contribute to this residual. In particular, electrical measurements identified a finite DC input current in the SP2 threshold-input stage. The three threshold references,
$V_{\mathrm{th0}}$, $V_{\mathrm{th1}}$, and $V_{\mathrm{th2}}$, are routed
through a shared bias infrastructure serving the two superpixel variants.
Currents ranging from approximately $100~\mathrm{nA}$ to $3~\mu\mathrm{A}$
were measured on these lines, depending on threshold setting and operating
condition. These currents arise from the finite input impedance of the SP2
threshold interface and should not be interpreted as semiconductor-junction
leakage.

The measured current can produce voltage drops in the resistive threshold
distribution network and modify the effective discriminator biases. Circuit
simulations confirm the excessive DC loading of the SP2 threshold-input
stage and predict substantially smaller input currents for SP1. Because part
of the upstream threshold-bias infrastructure is shared, the SP2 loading can
also perturb the voltages supplied to SP1. This mechanism is therefore a
plausible contributor to the residual threshold dispersion.

The pixel-level threshold maps, however, do not exhibit a sufficiently
reproducible monotonic spatial pattern to determine the magnitude of this
contribution directly. The remaining dispersion may also include residual
charge-injection nonuniformity, local baseline variation, and mismatch not
fully represented in the pre-layout simulations. The measured
78.24~\unit{\electron} value should therefore be interpreted as an upper bound on the
intrinsic comparator-threshold dispersion rather than as a direct
measurement of residual auto-zero mismatch alone.

\subsection{Design Implications and Next Steps}

The first-silicon measurements validate the intrinsic noise performance of the
AFE while identifying several implementation-specific limitations.
Of the two variants, SP1 provides the more uniform and predictable operating
point under the conditions investigated in this work. Consequently, all
subsequent system-level measurements, including validation of the on-chip
neural-network classifier, were performed using SP1 and the operating point
established in the preceding subsections.

The characterization results motivate the following modifications for the next
prototype revision:

\begin{itemize}
    \item \textbf{Calibration network:} Replace the shared global injection line
    with local in-pixel charge injection. This modification will substantially
    reduce the aggregate capacitive loading, pulse-shape distortion, and
    position-dependent calibration errors observed when multiple pixels are
    excited simultaneously.

    \item \textbf{Low-threshold linearity:} Reduce the sampling-switch charge
    injection responsible for the approximately 50~\unit{\electron} effective offset at
    low discriminator thresholds. Increasing the auto-zero capacitance by a
    factor of ten is expected to reduce the voltage perturbation produced during
    sampling and extend the linear operating range toward the intended
    400~\unit{\electron} minimum threshold.

    \item \textbf{Threshold-bias distribution:} Redesign the SP2 threshold-input
    stage to present a high DC input impedance to
    $V_{\mathrm{th0}}$, $V_{\mathrm{th1}}$, and $V_{\mathrm{th2}}$.
    The finite input impedance of the present implementation draws current from
    the shared threshold grid and produces routing-dependent voltage drops that
    affect both superpixel variants. Buffering or separating the threshold
    distribution networks will further reduce this systematic contribution to
    the measured threshold dispersion.
\end{itemize}

Despite these first-silicon non-idealities, the measured SP1 ENC of approximately 55~\unit{\electron} agrees closely with circuit simulations for the front end without an attached sensor. The prototype therefore provides a stable and reproducible platform for evaluating the complete mixed-signal processing chain. The planned design modifications are intended to preserve this validated noise performance while improving threshold uniformity, calibration accuracy, and operating margin at the target minimum threshold of 400~\unit{\electron}.

As a next step, operation at the nominal 40~MHz bunch-crossing frequency must be validated experimentally. As noted in Sec.~\ref{sec:setup} the \texttt{CaR} board is limited in bandwidth, consequently \texttt{Fermilab} has developed an upgraded interface capable of providing higher-bandwidth digital signals for operation at and beyond the nominal 40~MHz frequency. Future measurement campaigns will evaluate operation at the nominal 40~MHz
bunch-crossing frequency and at temperatures down to approximately
$-30\,^{\circ}\mathrm{C}$. These studies will determine the bias current required
to satisfy the higher-bandwidth operating condition and quantify the temperature
dependence of the ENC, threshold dispersion, and threshold-bias current.

\section{Digital On-Chip NN Characterization}
\label{sec:dnn_results}

Following AFE validation, calibrated charge profiles corresponding to realistic particle clusters were injected into the pixel matrix. The digitized pixel outputs were summed along the local column direction (defined as the `$y$-profile', see Fig.~\ref{fig:ROICpix_b}) and subsequently sent as input to the on-chip NN. Due to its combinational nature, the NN output evolves dynamically as signal transitions propagate through the inference logic. Capturing the final classification state off chip requires sufficient temporal resolution. The FPGA therefore samples the NN output every 2.5~ns over an observation window spanning 1.5 \texttt{BxCLK} periods, and the resulting time-resolved data are stored for offline analysis. 

The NN weights and biases were programmed through the DAQ system, and both the $y$-profile inputs and NN classification outputs were recorded on every clock cycle. The NN class output encodes three particle categories: high-$p_{\mathrm{T}}$, low-$p_{\mathrm{T}}$ positive charge, and low-$p_{\mathrm{T}}$ negative charge.

\subsection{Digital Power Measurements}

One advantage of the combinational design of the NN is that it requires neither clock distribution nor sequential storage elements within the inference datapath. Consequently, dynamic power is consumed only when the input $y$-profile changes. Such transitions occur only during particle interactions or when stochastic noise induces a change in a pixel’s digital output. When the input remains unchanged, the logic is electrically static and consumes negligible dynamic power.

The power consumption of the on-chip NN was evaluated under conditions representative of the inner tracking detectors of the CMS experiment at the HL-LHC. Charge clusters were injected at a rate corresponding to the expected particle occupancy, which, when scaled to the 1.6 mm$^2$ active area of the prototype, corresponds to an average of one cluster every 1.56~\unit{\micro\second}. The [V$_\mathrm{th0}$, V$_\mathrm{th1}$, V$_\mathrm{th2}$] values were chosen to be [1000, 1600, 2400]~\unit{\electron} to perform this measurement since these were found to be the nominal values for best on-chip NN performance.

\begin{figure}
    \centering
    \includegraphics[width=0.5\linewidth]{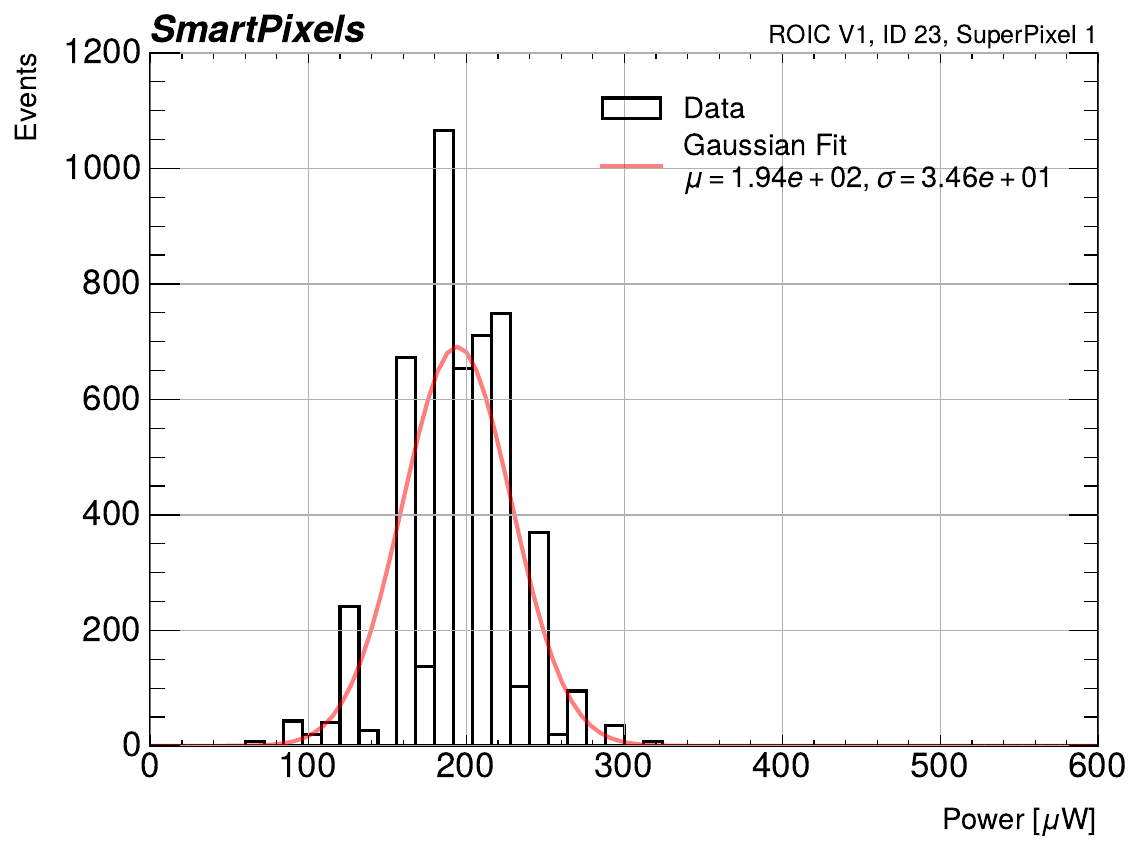}
    \caption{Histogram containing values of the power consumed by the on-chip NN when charge clusters were injected at a rate corresponding to the expected particle occupancy. Here, each event corresponds to charge clusters injected at a rate similar to expected rate at HL-LHC conditions in the CMS experiment.}
    \label{fig:power-measurements}
\end{figure}

The obtained result is shown in Fig.~\ref{fig:power-measurements}, \textbf{where from a sample of $10^4$ injected clusters an average power consumption of 194 $\mu$W was measured for the full 256-pixel array, corresponding to 0.76 $\mu$W per pixel}. The observed spread in power consumption reflects variations in cluster size, inferred class, and noise-driven switching activity, which together determine the active signal paths within the combinational logic.

\subsection{On-Chip Versus RTL Simulation Agreement}

The $y$-profiles that are sent as input to the on-chip NN are saved using the scan-chain, and are subsequently used to obtain the expected NN output using offline RTL simulations. We studied the on-chip NN under varying noise occupancies by measuring the agreement between the on-chip NN output and offline RTL simulations as a function of the comparator thresholds (V$_{\mathrm{th0}}$). The chosen range of V$_{\mathrm{th0}}$ values spanned from 400 to 1000~\unit{\electron}. The 400~\unit{\electron} minimum reflects the previously noted (Section~\ref{sec:qth_enc}) baseline for optimal signal-to-noise performance. At the upper bound of 1000~\unit{\electron}, noise-induced transitions are strongly suppressed, rendering results from higher threshold values redundant.

\begin{figure}[t!]
    \centering
    \includegraphics[width=0.5\linewidth]{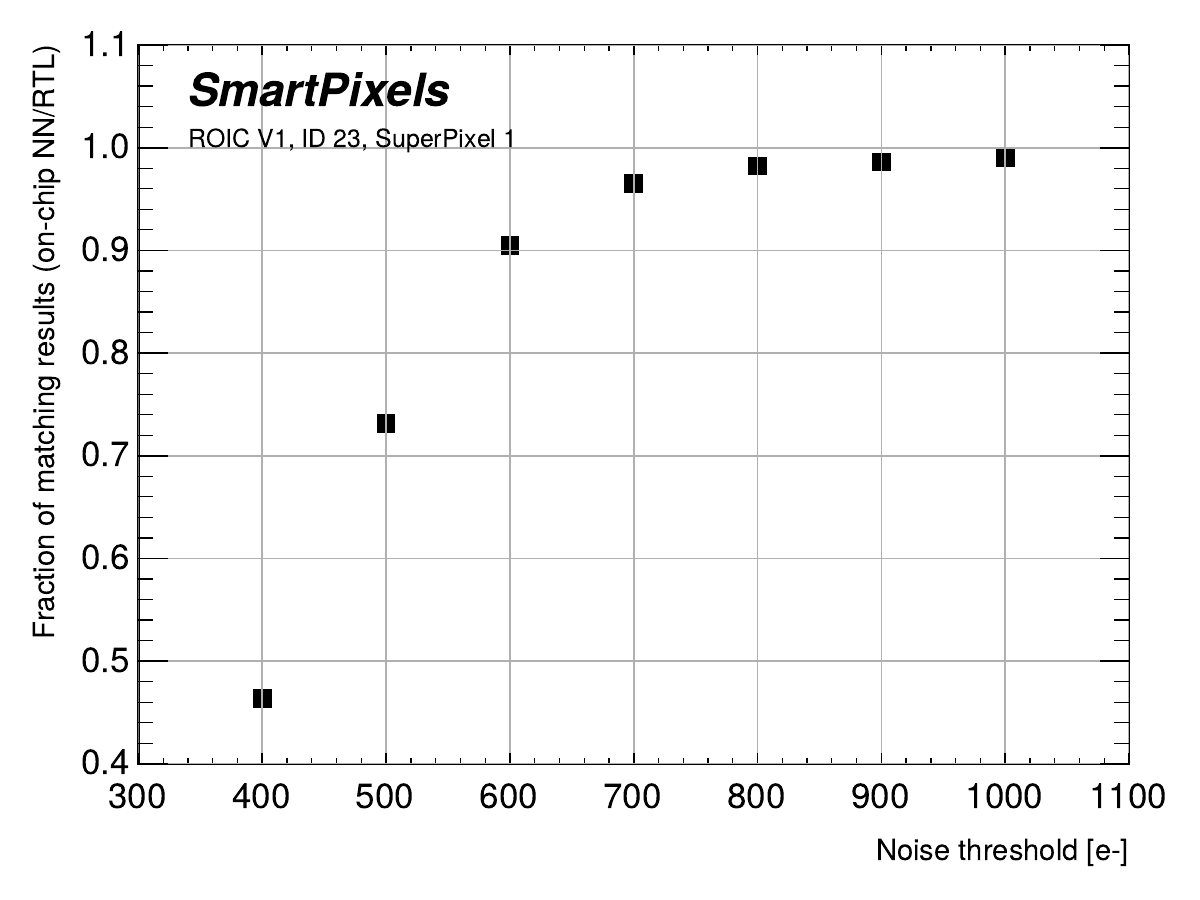}
    \caption{Measured fraction of events for which the on-chip NN classification matches the RTL simulation as a function of the effective charge threshold.}
    \label{fig:power_rtl_vs_noise}
\end{figure}

Fig.~\ref{fig:power_rtl_vs_noise} shows the fraction of events for which the on-chip NN classification matches the RTL simulation as a function of the noise threshold (V$_{\mathrm{th0}}$).  At low threshold values, the agreement is poor because AFE noise produces rapid spurious transitions after digitized $y$-profile is readout and because the off-chip system cannot reliably capture the resulting high-speed NN output.

\begin{figure}[t!]
    \centering
    \includegraphics[width=0.5\linewidth]{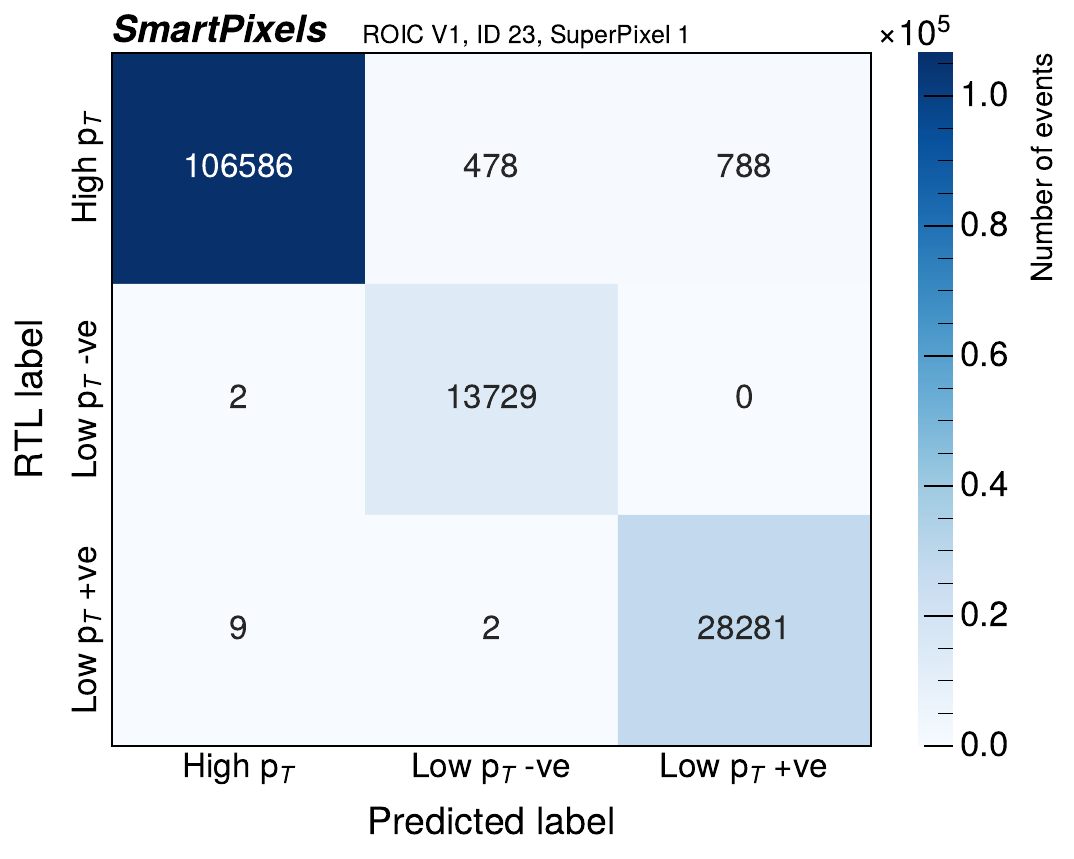}
    \caption{Confusion matrix for the on-chip NN at the 1000~\unit{\electron} threshold. The X and Y-axes represent the predicted class from the on-chip NN and offline RTL simulation, respectively, and the color axis represents the number of events.}
    \label{fig:nn_confusion_matrix}
\end{figure}

Future ASIC revisions will incorporate dedicated on-chip output-capture and readout circuitry to eliminate this measurement limitation. As the threshold is increased, these effects are progressively suppressed, and the agreement rapidly approaches 100\%. \textbf{For threshold settings corresponding to injected charges of approximately 1000~\unit{\electron} and above, the measured RTL–silicon agreement exceeds 99.06\%.} A representative confusion matrix for the 1000~\unit{\electron} threshold is shown in Fig.~\ref{fig:nn_confusion_matrix}; its strong diagonal structure confirms bit-accurate agreement between the on-chip NN and the RTL simulation.

These results demonstrate that the digital logic operates deterministically in the prototype’s stable analog regime. At lower thresholds, the mismatches are dominated by noise-related effects rather than intrinsic differences between the synthesized RTL and the silicon implementation. A detailed study on the classification performance of the on-chip neural network will be presented in a forthcoming paper. Future iterations of the ASIC will target AFE and readout enhancements to enable robust operation at thresholds comparable to those used in HL-LHC pixel detectors. In particular, we are interested in improving performance at low threshold values as it increases the chances for small collected charges to still register as hits, especially after irradiation when charge collection degrades and noise rises. Consequently, we also expect improvements in the performance of the network on-chip due to the enhanced cluster information.

\section{Conclusion and Outlook}
\label{sec:conclusion}

This work presents the design and first-silicon characterization of a 28-nm CMOS pixel readout ASIC that integrates synchronous charge
digitization and on-chip neural network inference within a $25\times25~\mu\mathrm{m}^{2}$ pixel architecture. The prototype contains two $32\times8$ superpixels implementing differential and single-ended analog front-end variants, together with a fully combinational digital
processing chain for cluster-based filtering. The measurements exercise the complete signal path from calibrated charge injection through analog
amplification, multi-level digitization, feature formation, and on-chip classification. At the selected 10~MHz operating point, the analog front
end dissipates approximately 2.8~$\mu$W per pixel, while the neural-network logic consumes approximately 0.76~$\mu$W per pixel under a cluster rate representative of the expected HL-LHC occupancy. The resulting combined analog and digital processing power is approximately 3.6~$\mu$W per pixel, corresponding to an active-matrix power density of 0.57~W/cm$^{2}$. At the nominal 40~MHz operating frequency, the analog power is estimated to increase to approximately 5~$\mu$W per pixel. The digital power is expected to remain unchanged since it depends on hit density and not on the \texttt{BxCLK} frequency. The resulting total power is estimated at 5.76~$\mu$W per pixel, or  0.92~W/cm$^{2}$, which remains within the 1~W/cm$^{2}$ system-level power-density target.

At room temperature and a \texttt{BxCLK} frequency of 10~MHz, SP1 achieves a mean conversion gain of $\sim$57.7~\unit{\micro\volt\per\electron}, an average ENC of $\sim$54.6~\unit{\electron}, and a threshold dispersion of $\sim$78.2~\unit{\electron}. The measured ENC closely matches circuit simulations for the front end without an attached sensor, validating the intrinsic noise performance of the fabricated CSA. The characterization also revealed prototype-specific limitations associated with the shared calibration network, sampling-switch charge injection at low
thresholds, and DC loading of the common threshold-bias network. These effects degrade calibration accuracy and threshold uniformity but are implementation specific and do not represent fundamental limitations of the underlying front-end architecture.

The next ASIC revision will directly address these limitations by replacing the shared calibration line with local in-pixel charge injection, increasing the auto-zero capacitance to improve low-threshold linearity, and eliminating any DC loading of the shared threshold references. The digital readout will also incorporate on-chip capture of the neural-network output to remove limitations associated with high-speed off-chip sampling. Additional characterization is planned at the nominal 40~MHz bunch-crossing frequency, and at temperatures down to approximately -30$^{\circ}\mathrm{C}$.

Beyond the analog characterization, the on-chip neural network was validated by comparing its measured output with bit-accurate RTL inference. At effective discriminator thresholds of $\sim$1000~\unit{\electron} and above, the RTL-to-silicon agreement exceeds 99\%, demonstrating deterministic event-by-event operation of the implemented digital logic within the stable analog operating regime. This comparison validates the silicon implementation of the classifier, but the developments needed to improve and validate the on-chip network performance in realistic detector conditions remain ongoing.

In parallel, ongoing work is investigating noise-aware retraining of the neural network using charge profiles augmented with the measured ENC,
pixel-to-pixel threshold dispersion, and baseline fluctuations. This study will determine how much of the low-threshold classification performance can be recovered through algorithmic robustness while the planned hardware modifications improve the analog operating margin. It will also quantify the tradeoffs among discriminator threshold, noise occupancy, data reduction, and classification accuracy, while evaluating the extent to which noise-aware retraining can recover performance degraded by severe radiation damage under realistic detector conditions. Together, these hardware and algorithmic developments provide a path toward reliable intelligent data filtering at the intended 400~\unit{\electron} operating threshold and at the 40~MHz bunch-crossing rate.

\label{sec:conclusion}

\section*{Acknowledgements}

This work was completed using computing resources at the University of Chicago’s Research Computing Center and the Fermilab Elastic Analysis Facility. 
%
%
We would like to extend our sincere gratitude to Harish Jamakhandi and David Burnette from Siemens EDA for their assistance and expertise with Catapult HLS. 
%
DB, GDG, FF, AG, LG, JH, RL, BP, and CS are supported by Fermi Forward Discovery Group, LLC under Contract No. 89243024CSC000002 with the U.S. Department of Energy, Office of Science, Office of High Energy Physics. GDG and LG are partially supported under DOE project "HAAI: Designing Smart Detectors with a ML-to-Silicon Platform" (LAB 24-3305). GDG and BP are also supported by the U.S. Department of Energy, Office of Science, Offices of High Energy Physics and Advanced Scientific Computing Research, as part of the Co-design and Heterogeneous Integration in Microelectronics for Extreme Environments (CHIME) Microelectronics Science Research Center (MSRC) (DOE Lab Announcement No. LAB 24-3320).
%
JD and BW are supported by the DOE Early Career Research Program award DE-SC0026236.
NT is also supported by the DOE Office of Science, Office of Advanced Scientific Computing Research under the “Real-time Data Reduction Codesign at the Extreme Edge for Science” Project (DE-FOA-0002501).
%
MS is supported by NSF-PHY award 2012584. 
%
DS is supported by the Visiting Scholars Award Program of the Universities Research Association. CM and DS are supported by NSF-PHY award 2513237. ART and CM are supported by DOE DE-SC0023715 (482662 / EB645) ``pixel-integrated neural processing for extreme-precision collision-event readout''.
%
KFD and EH are supported by the NSF CAREER Program through award 2443370, and KFD is additionally supported by the Neubauer Family Assistant Professor Program. EY, AN, and DA are supported by the University of Chicago’s Quad Undergraduate Research Scholar program, the Jeff Metcalf Internship program, and the Sachs Fellowship, respectively.
%
MSN is supported through NSF cooperative agreement OAC-2117997 and the DOE Office of Science, Office of High Energy Physics, under Contract No. DE-SC0023365.
%
A. Badea is supported by the Schmidt Sciences Foundation.
\bibliographystyle{JHEP}
\bibliography{biblio}

@article{CMS_2008,
author= {{CMS Collaboration}},
doi = {10.1088/1748-0221/3/08/S08004},
url = {https://dx.doi.org/10.1088/1748-0221/3/08/S08004},
year = {2008},
month = {aug},
publisher = {},
volume = {3},
number = {08},
pages = {S08004},
title = {The CMS experiment at the CERN LHC},
journal = {Journal of Instrumentation},
}

@article{Yoo_2024,
doi = {10.1088/2632-2153/ad6a00},
url = {https://doi.org/10.1088/2632-2153/ad6a00},
year = {2024},
month = {aug},
publisher = {IOP Publishing},
volume = {5},
number = {3},
pages = {035047},
author = {Yoo, Jieun and Dickinson, Jennet and Swartz, Morris and Di Guglielmo, Giuseppe and Bean, Alice and Berry, Douglas and Blanco Valentin, Manuel and DiPetrillo, Karri and Fahim, Farah and Gray, Lindsey and Hirschauer, James and Kulkarni, Shruti R and Lipton, Ron and Maksimovic, Petar and Mills, Corrinne and Neubauer, Mark S and Parpillon, Benjamin and Pradhan, Gauri and Syal, Chinar and Tran, Nhan and Wen, Dahai and Young, Aaron},
title = {Smart pixel sensors: towards on-sensor filtering of pixel clusters with deep learning},
journal = {Machine Learning: Science and Technology},
}

@article{shekar2025,
title = {Robustness of the smartpixels classifier for different simulated sensor geometries and non-ideal detector conditions},
journal = {Nuclear Instruments and Methods in Physics Research Section A: Accelerators, Spectrometers, Detectors and Associated Equipment},
volume = {1092},
pages = {171897},
year = {2026},
issn = {0168-9002},
doi = {https://doi.org/10.1016/j.nima.2026.171897},
url = {https://www.sciencedirect.com/science/article/pii/S0168900226006236},
author = {Danush Shekar and Ben Weiss and Morris Swartz and Corrinne Mills and Jennet Dickinson and Lindsey Gray and David Jiang and Mohammad Abrar Wadud and Daniel Abadjiev and Anthony Badea and Douglas Berry and Alec Cauper and Arghya Ranjan Das and Karri Folan DiPetrillo and Farah Fahim and Rachel {Kovach Fuentes} and Abhijith Gandrakota and Giuseppe {Di Guglielmo} and Eliza Howard and Shiqi Kuang and Carissa Kumar and Mia Liu and Petar Maksimovic and Nick Manganelli and Mark S. Neubauer and Aidan Nicholas and Emily Pan and Benjamin Parpillon and Jannicke Pearkes and Ricardo Silvestre and Chinar Syal and Amit Trivedi and Keith Ulmer and Jieun Yoo and Eric You},
}

@misc{smartpix_regression,
      title={On-chip probabilistic inference for charged-particle tracking at the sensor edge}, 
      author={Arghya Ranjan Das and David Jiang and Rachel Kovach-Fuentes and Shiqi Kuang and Ana Sofía Calle Muñoz and Danush Shekar and Jennet Dickinson and Giuseppe Di Guglielmo and Lindsey Gray and Mia Liu and Corrinne Mills and Mark S. Neubauer and Daniel Abadjiev and Anthony Badea and Doug Berry and Karri DiPetrillo and Farah Fahim and Abhijith Gandrakota and Harshul Gupta and James Hirschauer and Eliza Howard and Ron Lipton and Petar Maksimovic and Nick Manganelli and Benjamin Parpillon and Jannicke Pearkes and Ricardo Silvestre and Morris Swartz and Chinar Syal and Nhan Tran and Amit Trivedi and Keith Ulmer and Mohammad Abrar Wadud and Benjamin Weiss and Eric You},
      year={2026},
      eprint={2602.15946},
      archivePrefix={arXiv},
      primaryClass={physics.ins-det},
      url={https://arxiv.org/abs/2602.15946}, 
}

@misc{zenodo_16x16,
  author       = {Shekar, Danush and
                  Swartz, Morris and
                  Dickinson, Jennet and
                  Wadud, Mohammad Abrar},
  title        = {Smartpixels 16x16 datasets},
  month        = feb,
  year         = 2026,
  publisher    = {Zenodo},
  version      = 1,
  doi          = {10.5281/zenodo.18472791},
  howpublished = {https://doi.org/10.5281/zenodo.18472791},
}

@article{ISCAS2023,
     author  = "Benjamin Parpillon and Amit Trivedi and Farah Fahim", 
     title   ="{Readout IC with 40 MSPS in-pixel ADC for future vertex detector upgrades of Large Hadron Collider}",
     year    = "2023",
     journal = "2023 IEEE International Symposium on Circuits and Systems (ISCAS)",
     doi = "10.1109/ISCAS46773.2023.10182033",
     url           = "https://ieeexplore.ieee.org/document/10182033",

}

@inproceedings{Quinn:2024xhl,
    author = "Quinn, Adam",
    title = "{An Open-Source Framework for Rapid Validation of Scientific ASICs}",
    eprint = "2406.15181",
    archivePrefix = "arXiv",
    primaryClass = "physics.ins-det",
    reportNumber = "FERMILAB-CONF-24-0206-ETD",
    month = "6",
    year = "2024"
}

@article{Vanat:2703500,
      author        = "Vanat, Tomas",
      title         = "{Caribou - A versatile data acquisition system}",
      reportNumber  = "CLICdp-Conf-2019-012",
      journal       = "PoS",
      volume        = "TWEPP2019",
      pages         = "100",
      year          = "2020",
      url           = "https://cds.cern.ch/record/2703500",
      doi           = "10.22323/1.370.0100",
}

@unpublished{Parpillon:2024TWEPP,
  author       = {Ben Parpillon},
  title        = {Radiation-Hard Smart-Pixel Detector ASIC ReadOut with Digital AI in 28\,nm},
  note         = {Poster presentation at TWEPP 2024, Topical Workshop on Electronics for Particle Physics, Glasgow, United Kingdom, 30 September--4 October 2024},
  howpublished = {\url{https://indico.cern.ch/event/1381495/contributions/5988566/}},
  year         = {2024},
}

@INPROCEEDINGS{parpillon2024smartpixelsinpixelai,
  author={Parpillon, B. and Syal, C. and Yoo, J. and Swartz, M. and Guglielmo, G. Di and Bean, A. and Berry, D. and Valentin, M. Blanco and DiPetrillo, K. and Badea, A. and Gray, L. and Maksimovic, P. and Mills, C. and Neubauer, M. and Pradhan, G. and Tran, N. and Wen, D. and Fahim, F.},
  booktitle={2024 IEEE Nuclear Science Symposium (NSS), Medical Imaging Conference (MIC) and Room Temperature Semiconductor Detector Conference (RTSD)}, 
  title={Smart Pixels: In-pixel AI for on-sensor data filtering}, 
  year={2024},
  volume={},
  number={},
  pages={1-2},
  doi={10.1109/NSS/MIC/RTSD57108.2024.10655003}}

@online{smartpixels_website,
  author  = {{Fast Machine Learning}},
  title   = {{Smart Pixels}},
  url     = {https://fastmachinelearning.org/smart-pixels/},
  urldate = {2026-08-11}
}

@misc{RD53BmeasGaioni,
  title = {CMS analog front-end: simulations and measurements},
  howpublished = {\url{https://cds.cern.ch/record/2746420/files/Linear%20AFE%20report.pdf}},
  author = {Luigi Gaioni},
  urldate = {2021-10-20}
}

@misc{RD53Bspec,
  title = {RD53B users guide: Introduction to RD53B pixel chip architecture, features and recommendations for use in pixel detector systems},
  howpublished = {\url{https://cds.cern.ch/record/2754251/files/users_guide_v11.pdf}},
  author = {RD53 collaboration},
  urldate = {2021-05-10}
}

@article{ITKv1,
  title={Novel 3D Pixel Sensors for the Upgrade of the ATLAS Inner Tracker},
  author={Stefano Terzo,Maurizio Boscardin, Et Al.},
  journal={Frontiers in Physics},
  volume={9},
  year={2021},
}

@article{Crocv1,
  title={Analog performance of the CROCv1 pixel readout chip for the CMS Phase-2 Tracker Upgrade},
  author={A. Papadopoulos and on the CMS Tracker Group},
  journal={Nuclear Instruments and Methods in Physics Research Section A: Accelerators, Spectrometers, Detectors and Associated Equipment},
  volume={18},
  year={2023},
  publisher={Journal of Instrumentation}
}

\end{document}